\documentstyle[11pt,newpasp,twoside,epsf]{article}

\newcommand{\asca}{{\it ASCA} }
\newcommand{\chandra}{{\it Chandra} }
\newcommand{\hetg}{{\it HETG} }

\newcommand{\etal}{{\it et al.} }

\def\edcomment#1{\iffalse\marginpar{\raggedright\sl#1\/}\else\relax\fi}
\reversemarginpar

\begin{document}

\title{\chandra Grating Observations of Seyfert 1 Galaxies}
\author{T. Yaqoob}
\affil{Johns Hopkins University, 3400 N. Charles St., Baltimore, MD21218. NASA/GSFC, Code 662, Greenbelt Rd., Greenbelt, MD20771.}

\author{I. M. George, T. J. Turner}
\affil{Joint Center for Astrophysics, University of Maryland, Baltimore County, 1000 Hilltop Circle, Baltimore, MD21250.
NASA/GSFC, Code 662, Greenbelt Rd., Greenbelt, MD20771.}

\begin{abstract}
We present new results from a \chandra \hetg
observation of NGC~5548 and give a comparison
of absorption and emission features found in 
Seyfert 1 galaxies using \chandra grating observations.
Deep soft X-ray edges are seen in Mkn~509 and NGC~3783,
consistent with \asca data. In NGC~5548 and NGC~4501 the
edges are weak but consistent with the low column densities. 
We show that the detection of a narrow, probably non-disk,
component of the iron line is very common. 
We show the effect of
removing this narrow component from the \asca Fe-K line profile
in NGC~4151, revealing
the underlying true shape of the relativistic Fe-K line component.
\end{abstract}

\section{HETG Observation of NGC~5548}

Our preliminary \hetg spectra of NGC~5548
(exposure time $\sim 80$ ks) taken in 2000, February,
are shown in Fig. 1 and Fig. 2(a) (see also George, Turner, Yaqoob 2001). 
The strongest features (apart from Fe~K$\alpha$ -- see Yaqoob \etal 2001) --
are the O{\sc vii} He-like triplet and absorption features due
to Ne and O $Ly\alpha$, and 
O{\sc vii}~$1s^{2}-3p \ ^{1}P_{1}$. See Table 1 for a comparison with 
Kaastra \etal (2000). O{\sc vii} (f) is particularly strong compared to
(i) and (r), indicating a photoionization-dominated plasma. 
It is consistent with being at the systemic velocity but we measure
a range of blueshifts for the absorption lines. These, and detailed
analysis will be reported in 
McKernan \etal (in preparation).

\begin{figure}[thb]
\plotfiddle{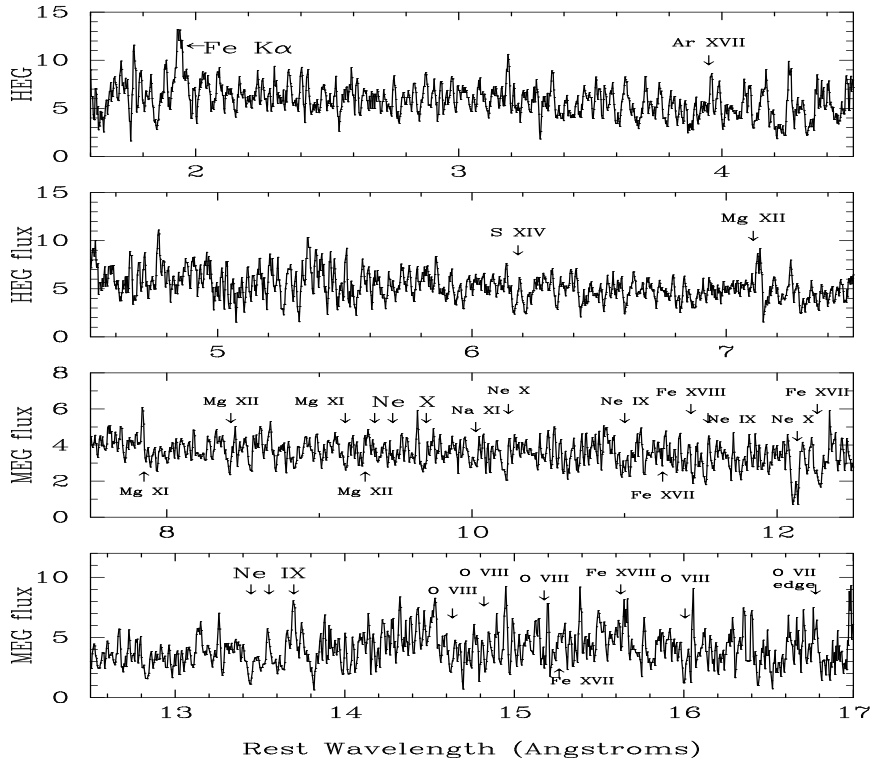}{4in}{0.0}{80}{45}{-220.0}{-10.0}
\caption{HEG and MEG spectra of NGC~5548. Note that line IDs are simply
marked for convenience and do not necessarily indicate significant detections.
}\label{n5548big}
\end{figure}


\section{Comparison of Soft X-ray Absorption and Emission Features}

\begin{figure}[tbh]
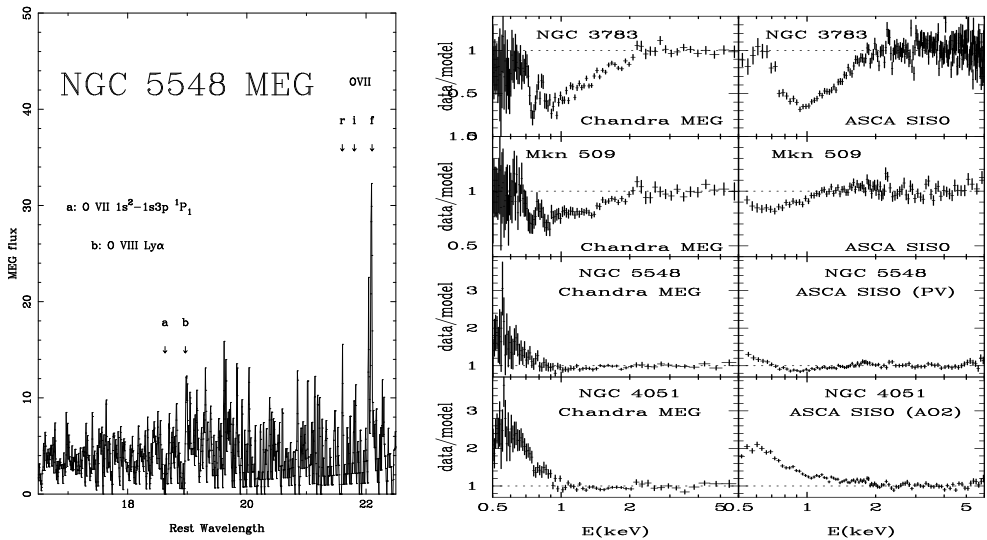

\vspace{-12cm}
\plotfiddle{m_ty_fig2.ps}{4in}{0.0}{30}{30}{-200.0}{-300.0}
\plotfiddle{m_ty_fig3.ps}{4in}{0.0}{55}{30}{-80.0}{0.0}
\caption{{\bf (a) Left:} MEG spectra of NGC~5548 continued.
{\bf (b) Right:} \chandra MEG and (non-contemporaneous) \asca SIS  spectra of
four Sy~1. The 2--5 keV data were independently fitted with a power law,
which was extrapolated to low energies. 
}\label{n5548sml}
\end{figure}

Detailed \chandra grating spectroscopy has now been published
for five Seyfert 1 galaxies, as summarized in Table 1.
Kaspi \etal (2001a) present the highest S/N grating data
available for any AGN ($\sim 900$ ks on NGC~3783; Table 1 refers only
to the shorter 53 ks observation). NGC~3227 was too weak for
searching for narrow absorption/emission features;
3C~273 is bright but no absorption/emission features
were detected (see George \etal 2001).
Ton~S~180 (Turner \etal 2001) and PG~1404 (Nandra \etal
in preparation), both NLS1, show smooth soft X-ray continua with
no narrow absorption/emission features.
Our own analysis of archival data for Akn~564, another NLS1, shows
a prominent, broad absorption feature at $\sim 0.7$ keV, resolved
by the MEG.
 
In addition to NGC~5548, at least two other sources show
multiple-velocity absorber systems
(NGC~4051, Collinge \etal 2001; NGC~3783, Kaspi \etal 2001b, George \etal 2001).
In all cases when absorption lines are detected they appear to be in
outflow, except in MCG~$-$6$-$30$-$15 (in which the lines
are consistent with the galaxy systemic velocity). In all cases where
emission lines are detected they appear to be consistent with the
galaxy systemic velocity, except for NGC~4151, in which the emission
is consistent with the small outflow velocity of the optical NLR.
See Table 1 for details.

\section{Soft X-ray Absorption Edges}

Fig. 2(b) compares the \chandra MEG soft X-ray spectra,
(binned at approximately \asca resolution), of four AGN
(NGC~3783, Mkn~509, NGC~5548, NGC~4051), with historical
\asca data. Given that these sources are time-variable,
the AGN show similar features with \chandra and {\it ASCA}.
Deep absorption edges  are seen in NGC~3783 and  Mkn~509.
These could be due to O{\sc vii} and O{\sc viii}, but
the lower energy edge could also be due to Fe{\sc I}
(locked up in dust?) as suggested by Lee \etal (2001)
for MCG~$-$6$-$30$-$15. However,
NGC~5548 and NGC~4051 have prominent soft excesses. If
soft excess is from a larger region than the
warm absorber, it could mask
underlying absorption features. 
However, the depth of the edges depends on the 
physical conditions of the asborber, so the apparent
absence of an edge should not be alarming.

\section{Narrow Fe-K Line Components Resolved by Chandra (HETG)}

As has been suspected for a long time the Fe-K lines in AGN are
generally complex, being composed of a narrow component (originating
far from the putative central black hole) and a broad component,
originating from closer to the black hole (e.g. see Weaver \etal 1997;
Yaqoob \etal 2001; Reeves \etal 2001). In a given observation 
of a given AGN only one of these may dominate, or they may both be comparable.
Therefore it is important to account for when
modeling the broad, relativistic disk lines. Some preliminary results
for \hetg measurements of the narrow line
are given in Table 2. Four of these \hetg narrow Fe-K lines are
shown in Fig. 3, along with the overall, total Fe-K lines measured
with non-contemporaneous \asca observations.
{\it 
These are single-Gaussian fits and may be misleading if the overall
line profile is
complex.} In fact the
large EW for NGC~4151 may be due to inadequate modeling of the
broad-line component and/or the continuum being particularly weak
in this observation. Errors are 90\% for one interesting parameter.
We are extending this census of the Fe-K region using \chandra
\hetg (simultaneous with other missions whenever possible) to many
other sources. Full results for NGC~3516, Mkn~509,
NGC~4593, F~9, 3C~120, NGC~7314, and more, 
will be reported in the near future.

\begin{center}
{\bf Table 1: Soft X-ray Grating Spectroscopy: Principal Features } \\
{\it Note: Fe~K$\alpha$ line is not discussed in this table.}
\end{center}
 
\footnotesize
\begin{minipage}[b]{2.5cm}
\par\noindent
{\bf NGC~5548} \\
LETG/HRC \\
Kaastra \etal \\
(2000) \\
\\
 
{\bf NGC~5548} \\
HETG/ACIS \\
George \etal \\
(in prep.) \\
\\
\\
\\
 
{\bf NGC~3783} \\
HETG/ACIS \\
Kaspi \etal \\
(2001) \\
\\
\\
\\
\\
 
{\bf NGC~4051} \\
HETG/ACIS \\
Collinge \etal \\
 (2001) \\
\\
\\
 
{\bf MCG} \\
{\bf $-$6$-$30$-$15} \\
HETG/ACIS \\
Lee \etal \\
(2001) \\
\\
 
{\bf NGC~4151} \\
HETG/ACIS \\
Ogle \etal \\
(2001) \\
\\
\\
\end{minipage}
\begin{minipage}[b]{10cm}
\vspace{-2mm}
{\bf Strongest absorption features:} $Ly\alpha$ lines of
N, O, Ne, Mg. Resonance lines of He-like O, Ne. \\
{\bf Strongest emission features:} O{\sc VII} triplet;
forbidden line strongest. \\
{\bf Mean absorption-line velocity}: $-280 \pm 70$ km/s. \\
{\bf Mean emission-line velocity}: $200 \pm 130$. \\

{\bf Strongest absorption features:}
Ne $Ly\alpha$, O $Ly\alpha$, O{\sc VII}~$1s^{2}-3p \ ^{1}P_{1}$
(see Figs. 1 and 2(a)). \\
{\bf Strongest emission features:} O{\sc VII} triplet;
forbidden line strongest. \\
{\bf Mean absorption-line velocity}: difficult to measure but
consistent with LETG/HRC results. \\
{\bf Mean emission-line velocity}: consistent with galaxy systemic
velocity. \\

{\bf Strongest absorption features:}
Many He and H-like states of O, Ne, Mg (Si weaker).  Many states of
Fe. \\
{\bf Strongest emission features:} Several transitions of
He and H-like states of O, Ne, Mg (Si very weak).
The O{\sc VII} triplet is the strongest emission feature (of which
$f$ is strongest). \\
{\bf Mean absorption-line velocity}: $-610 \pm 130$ km/s. \\
{\bf Mean emission-line velocity}: consistent with galaxy systemic
velocity. \\
 
{\bf Strongest absorption features:} $Ly\alpha$ lines of
Ne, Mg, Si. \\
{\bf Strongest emission features:} He-like triplets of O, Ne.
$f$ strongest in both cases. \\
{\bf Mean absorption-line velocity}: Two systems; $-2340 \pm 130$ km/s
and $-600 \pm 130$ km/s.
{\bf Mean emission-line velocity}:
consistent with galaxy systemic velocity. \\
 
{\bf Strongest absorption features:}
 $Ly\alpha$ lines of N, O; Ne{\sc IX}~$1s^{2}-1s2p$; Fe L edge. \\
{\bf Strongest emission features:} No strong lines. \\
{\bf Mean absorption-line velocity}: $|v| <200$ km/s,
consistent with galaxy systemic velocity. \\
 
{\bf Strongest absorption features:} No obvious absorption features. \\
{\bf Strongest emission features:} Spectrum dominated by NLR emission.
Many lines; those with S/N $>5$ are $Ly\alpha$ of O, Ne, Mg;
$K\alpha$ of Si. $r$ and $f$ of He-like Mg. $f$ of He-like O, Ne. \\
{\bf Mean emission-line velocity}: $-120 \pm 50$ km/s,
consistent with velocity of optical NLR. \\
\\
\end{minipage}
\normalsize

\begin{figure}[t]
\plotfiddle{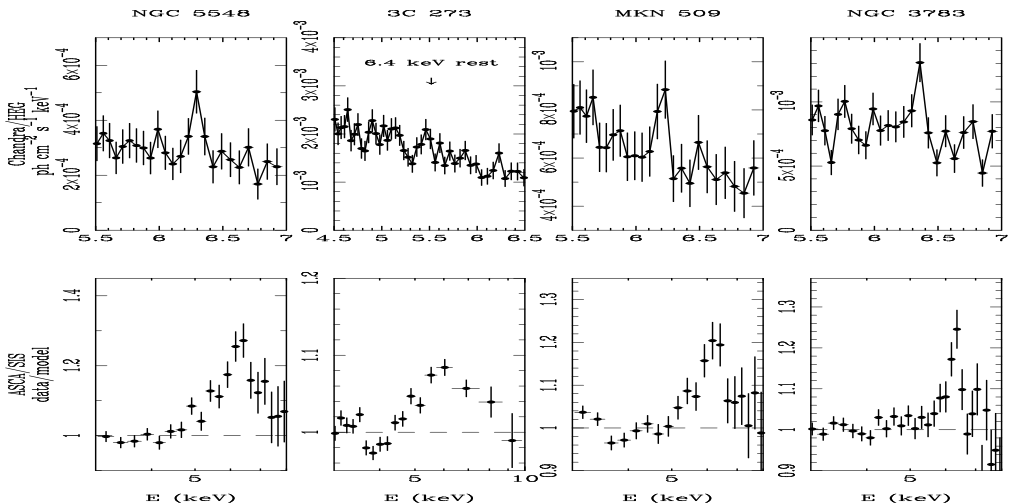}{4in}{0.0}{80}{30}{-230.0}{70.0}
\vspace{-3.5cm}
\caption{{\bf Top:} HEG spectra of the Fe-K region, showing the
detection of a narrow, non-relativistic component, in some AGN.
{\bf Bottom:} Non-contemporaneous \asca spectra in the Fe-K region showing
the full line profiles.} 
\label{fenarrow}
\end{figure}

\vspace{-4mm}
\setcounter{table}{1}
\begin{center}
\vspace{-10mm}
\begin{table}[thb]
\begin{center}
\caption{\bf {\large The Narrow Fe-K Line Component in Some AGN}}
\begin{tabular}{lccc}
\hline
Source & $E_{\rm Fe-K}$ (keV) & EW (eV) & FWHM (km/s) \\
\hline
NGC 5548 & $6.402 \ (+0.027,-0.025)$ & $133\ (+62,-54)$ & $<8040$ \\
NGC 3783 & $6.347\ (+0.050,-0.021)$ & $86 \ (+35,-28)$ & $<6665$ \\
NGC 4151 & $6.386 \ (+0.014,-0.016)$ & $330 \ (+63,-67)$ & $<8169$ \\
Mkn 509 & $6.389 \ (+0.082,-0.030)$ & $61 \ (+35,-21)$ & $<6290$ \\
NGC 4051 & $6.444 \ (+0.051,-0.052)$ & $198 (+99,-82)$ & $<8851$ \\
NGC 3327 & 6.4 (fixed) & $<85$ & - \\
3C 273 & 6.4 (fixed) & $<25$ & - \\
\hline
\hline
\end{tabular}
\end{center}
{\it Note: since completion of this work, reprocessing of the first HETG
observation data with CALDB 2.8 for 3C~273 has revealed a $\sim 3\sigma$ detection of a
narrow Fe-K line with EW $41(+24,-22)$ eV.}
\end{table}
\end{center}

\begin{figure}[h] 
\vspace{10pt}
\plotfiddle{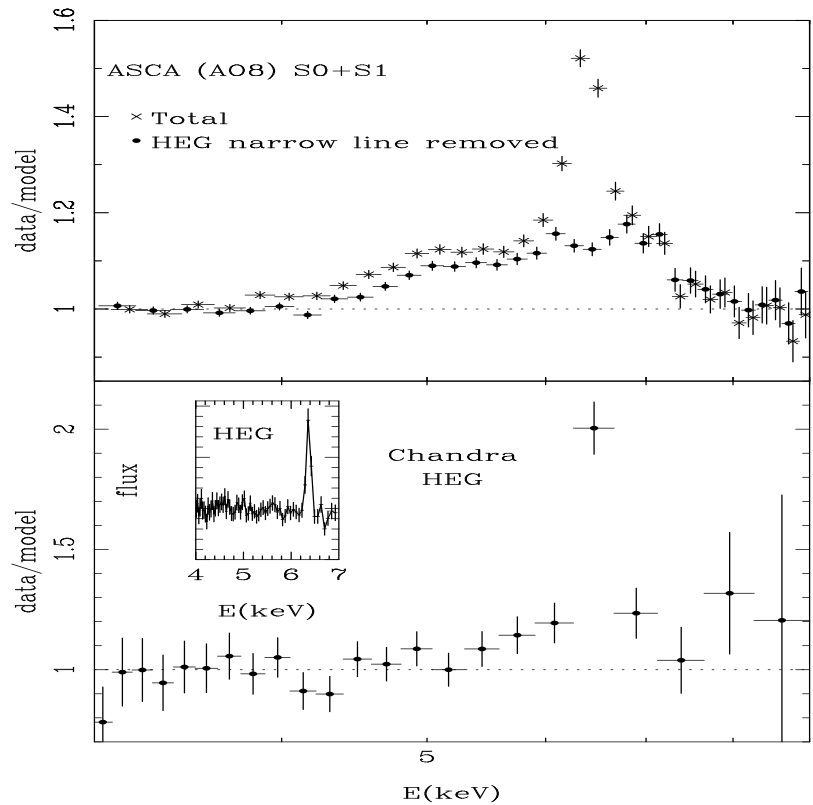}{4in}{0.0}{80}{45}{-200.0}{-20.0}
\caption{{\bf Top:} {\footnotesize The
true shape of the underlying broad Fe-K line in NGC~4151 (from the \asca AO8 SIS0$+$SIS1 data) when the
narrow line is modeled using the \chandra HETG parameters. {\bf Bottom:}
The Fe-K line in NGC~4151 observed by \chandra \hetg (the two profiles
correspond to the same data but binned by different factors, showing the
narrow and broad components of the line.}}\label{n4151line}
\end{figure}

Fig. 4  shows the true shape of the underlying broad
Fe-K line in NGC~4151 (from the \asca AO8 SIS data) when the
narrow line is modeled using the \chandra HETG parameters.
This would explain the puzzling result that,
if the narrow line is not taken into account, the total line profile
modeled with a relativistic disk
implies a near face-on disk, yet all other indications are
that the disk is viewed at much larger inclination angles in this
source.

{\bf Acknowledgments} We wish to thank B. McKernan, U. Padmanabhan, K. Weaver, Nandra, K.,  
for their help with parts of this work. We acknowledge support from the following grants: 
NCC5-447, NAS8-39073 (TY), NAG5-7067, NAG5-7385 (TJT).

\end{document}